\documentstyle[amsfonts,epsfig,preprint,aps]{revtex}

\begin{document}
\draft
\preprint{ }

\title{The ferromagnetic $q$-state Potts model on three-dimensional
lattices: a study for real values of $q$}

\author{S. Grollau\thanks{Corresponding author}, M.L. Rosinberg and G. Tarjus}\address{
Laboratoire    de   Physique    Th{\'e}orique   des Liquides\cite{AAAuth},
Universit{\'e} Pierre et Marie Curie,
\\ 4 place  Jussieu, 75252 Paris Cedex
05, France}
\maketitle

\begin{abstract}

We study the phase diagram of the  ferromagnetic $q$-state Potts model
on the various  three-dimensional lattices for integer and non-integer
values of  $q>1$.  Our  approach   is  based on  a   thermodynamically
self-consistent Ornstein-Zernike  approximation    for   the two-point
correlation functions.  We  calculate the transition temperatures and,
when the  transition is first  order, the  jump discontinuities in the
magnetization and the internal  energy, as well  as the coordinates of
the critical endpoint in external field.   Our predictions are in very
good agreement with  best available estimates.  From the numerical
study of the region of  weak first-order  transition, we estimate  the
critical value  $q_c$ for which the transition  changes from second to
first-order.   The $q\to 1^+$  limit  that describes the bond-percolation
problem is also investigated.

\end{abstract}

\

%\pacs{{\bf Pacs numbers}:05.50.+q, 75.10.H, 05.70.Jk}

\pacs{{\bf Key words}: Potts model, phase transitions, Ornstein-Zernike approximation.}

\def\be{\begin{equation}}
\def\ee{\end{equation}}
\def\bea{\begin{eqnarray}}
\def\eea{\end{eqnarray}}

\section{Introduction}\label{sec:introduction}

The Potts  model \cite{P52} is a  generalization of the Ising model to
more  (or less) than  two states per spin.   This model is realized in
many different   experimental situations  (see  Ref.~\cite{W82}  for a
review)    and it has been    the  subject of considerable theoretical
attention over the last two decades.  At low enough temperature and in
the absence of  a symmetry breaking  magnetic field, the ferromagnetic
Potts model undergoes a transition   from a disordered phase in  which
all  states  are   equally populated   to an   ordered  phase that  is
characterized by a non-zero  spontaneous magnetization associated with
preferential occupation of  one of the  $q$ states.  The order  of the
transition and the critical properties crucially depend on the spatial
dimension $d$  and the number of  states  $q$.  Whereas the mean-field
description  predicts the transition to  be  first-order for all $q>2$
irrespective of the dimension\cite{K54},  the  exact analysis  in  two
dimensions    shows    that   the   transition    is    continuous for
$q\leq4$\cite{Ba73}.  No exact results are available in three dimensions
but there is by  now general consensus that  the $3$-state Potts model
on the  simple cubic lattice undergoes  a weak  first-order transition
with   a  very    small   latent heat    and   a  large    correlation
length\cite{JV97}.  Since the  transition  is continuous for  $q\leq  2$
($q=2$  is  the standard   Ising  model whereas   the limit   $q\to  1^+$
corresponds   to the bond-percolation  problem),   the  order of   the
transition in 3-d must change at some $q_c$ between $2$ and $3$. There
are, however, only  few estimates of   $q_c$, and the behavior in  the
neighborhood of $q_c$ is not  precisely known\cite{NRS1981,KS82,LK91}.

To gain a  better perspective on   this issue and, more  generally, to
determine  accurately  the phase  diagram  of the  ferromagnetic Potts
model  on  the various   three-dimensional   lattices, we  propose  an
analytical treatment that allows us to  investigate integer as well as
non-integer  values   of $q$ ($q>1$).   Our  approach  derives from an
approximation that was first introduced to study the phase behavior of
simple  fluids\cite{HS77} and that  has been successfully applied to a
variety  of  spin  models,   including the   spin-$\frac{1}{2}$  Ising
model\cite{DS1996}, the Blume-Capel model\cite{GKRT2000} , and several
models of random  magnets\cite{KRT97}.  This approximation, called the
`` self-consistent  Ornstein-Zernike approximation'' (SCOZA), is based
on what  is   known  in  liquid-state statistical  mechanics    as the
Ornstein-Zernike  formalism, namely   (i)  the exact  Ornstein-Zernike
equations that relate the two-point connected correlation functions to
the two-point direct correlation functions (the former are obtained as
functional  derivatives  of the  field-dependent free-energy   and the
latter    as   functional derivatives    of   the Legendre-transformed
magnetization-dependent Gibbs  potential) and (ii) an Ornstein-Zernike
ansatz  that  assumes that the  direct  correlation functions have the
same range as  the interaction pair potential.   The dependence of the
correlation  functions   on the  control  variables  (temperature  and
magnetization) is determined by two types of constraints: on one hand,
exact  zero-separation   conditions   for the   connected  correlation
functions;  on the  other  hand, thermodynamic  self-consistency which
ensures that the same Gibbs free energy is obtained from the two-point
correlation functions, whether one integrates the internal energy with
respect to the inverse temperature or one integrates twice the inverse
susceptibility with respect to magnetization.

The general SCOZA   formalism  for the  ferromagnetic  $q$-state Potts
model in the presence of  $q-1$ external magnetic fields is  presented
in section~\ref{sec:presentation}.    In   section  III,  we  restrict
attention to the case of a  single symmetry breaking field that favors
one state while preserving the  symmetry  between the $q-1$  remaining
states.  This allows us  to study the thermodynamics  of the model for
any real  value of $q>1$  by  solving a  partial differential equation
with temperature  and  magnetization  as independent   variables.  The
numerical   results for the  various 3-d  lattices   are presented and
discussed   in section~\ref{sec:results}.  We  first   study the cases
$q=3$ and $q=4$ and compare to the  best available estimates.  We next
consider  $2<q<3$ and  provide  a numerical  estimate  of the critical
value  $q_c$ at  which the  transition  changes from  second to  first
order.  Finally,  the case $1<q\leq 2$ is discussed  and the limit $q\to 1^+$
is investigated.

\section{Self-consistent Ornstein-Zernike
approximation for the ferromagnetic Potts model}
\label{sec:presentation}

\subsection{Definitions and exact relations}

The $q$-state  Potts model  on a d-dimensional lattice of $N$ sites is
defined by the general Hamiltonian
\be
\label{Hamilt}
{\cal H}=-J \sum_{<ij>} (q{\delta}_{s_i,s_j}-1)-\sum_{i=1}^N \sum_{\alpha=1}^{q-1} h_i^{\alpha}(q\delta_{s_i,\alpha}-1)
\ee
where  $J$ is  a positive coupling  parameter  (we consider  here only
ferromagnetic  interactions),  $\delta_{s_is_j}$ is the  Kronecker symbol,
and the  sum  on $<ij>$ runs    over distinct nearest-neighbor  (n.n.)
pairs.   Each  site variable  $s_i$   can assume $q$  possible  values
$\alpha=1\ldots q$  and  the $h_i^{\alpha}$'s are   site-dependent magnetic fields
that  couple to    the  distinct states.   (Because  of  the condition
$\sum_{\alpha=1}^{q} \delta_{s_i,\alpha}=1$, only  $q-1$ fields are relevant and one
may take $h_i^q=0$ without loss  of generality.)  These site-dependent
fields have  been introduced   in order  to generate  the  correlation
functions (see below), but ultimately we are  interested in the system
in the presence of  a {\it uniform}  field that favors only one state.
The field-dependent free enegy ${\cal{F}}$ is then defined by

\be
\beta  {\cal F} =\tilde{\cal{F}}(\beta
J,\{\tilde{h}_i^{\alpha}\})=-  \ln Tr \exp[-\beta{\cal{H}}]
\ee
where    $\beta=1/(k_BT)$     is    the    inverse    temperature     and
$\tilde{h}_i^{\alpha}=\beta h_i^{\alpha}$.  The derivatives  of ${\cal F}$  with respect  to the
external  fields generate the local magnetizations $m_i^{\alpha}$,
\be
\label{mag}
m_i^{\alpha}=<\sigma_i^{\alpha}>=-\frac {\partial \tilde{\cal{F}}} {\partial
{\tilde{h}_i}^\alpha},  \hspace{1.2cm} \alpha=1,2,\ldots,(q-1)
\ee
 and the connected spin-spin correlation functions $G_{ij}^{\alpha\beta}$,
\be
\label{corG}
G_{ij}^{\alpha\beta}=<\sigma_i^{\alpha}\sigma_j^{\beta}>-<\sigma_i^{\alpha}> <\sigma_j^{\beta}>=- \frac {\partial^2 \tilde{\cal{F}}} {\partial
{\tilde{h}_i}^\alpha \partial {\tilde{h}_j^\beta }  } \hspace{1.2cm}  \alpha,\beta=1,2,\ldots,(q-1)
\ee
where     we      have    introduced       the     spin      variables
$\sigma_i^{\alpha}=q\delta_{s_i,\alpha}-1$.  Because  of the  condition $\sum_{\alpha=1}^{q}
\sigma_i^{\alpha}=0$,  there  are only $(q-1)$  independent magnetizations (per
site) and        $q(q-1)/2$    independent   correlation     functions
($G_{ij}^{\alpha\beta}= G_{ji}^{\beta\alpha}$) with

\be
m_i^q=-\sum_{\alpha=1}^{q-1}m_i^{\alpha}
\ee
and 
\be
G_{ij}^{\alpha q}=-\sum_{\beta=1}^{q-1} G_{ij}^{\alpha\beta}  \hspace{1.2cm} \alpha=1,2,\ldots,q.
\ee

Since we are interested in working with the magnetizations rather than
the fields as control parameters, we introduce  the Legendre
transformed Gibbs  potential 

\be
\label{Legendre}
\beta {\cal{G}}=\tilde{\cal{G}}(\beta J,\{m_i^\alpha\})
=\tilde{\cal{F}}+\sum_{i}\sum_{\alpha=1}^{q-1} m_i^\alpha  {\tilde{h}_i}^\alpha
\ee
which generates the direct correlation functions $C_{ij}^{\alpha\beta}$,

\be
\label{corC}
C_{ij}^{\alpha\beta} =\frac {\partial^2 \tilde{\cal{G}}} {\partial
m_i^\alpha \partial m_j^\beta   } \hspace{1.2cm} \alpha,\beta=1,2,...,(q-1).
\ee 
The two-point correlation functions $G_{ij}^{\alpha\beta}$ and the direct
correlation functions $C_{ij}^{\alpha\beta}$ are related by a set of 
Ornstein-Zernike (OZ) equations that result from the Legendre transform, 
\be
\label{OZ}
\sum_k\sum_{\gamma=1}^{q-1} C_{ik}^{\alpha\gamma}  G_{kj}^{\gamma\beta} =
\delta_{\alpha,\beta} \delta_{i,j} \hspace{1.2cm}  \alpha,\beta=1,2,...,(q-1).
\ee
In the limit of  uniform fields, one has $m_i^{\alpha}=m^{\alpha}$ and the two-point
correlation functions   depend only on the   vector ${\bf  r}$  that
connects the  two sites. The OZ equations  then take a  simple form in
Fourier space,
\be
\label{OZK}
\sum_{\gamma=1}^{q-1} \widehat{C}^{\alpha\gamma} ({\bf k})  \widehat{G}^{\gamma\beta} ({\bf
k}) =\delta_{\alpha,\beta} \hspace{1.2cm}  \alpha,\beta=1,2,...,(q-1) \ ,
\ee
and Eq. (8) yields the (inverse) susceptibility sum-rules:

\be
\label{ck}
\widehat{C}^{\alpha\beta}({\bf k=0}) =\frac {\partial^2 \tilde{\cal{G}}/N} {\partial
m^\alpha \partial m^\beta } \hspace{1.2cm} \alpha,\beta=1,2,...,(q-1) \ .
\ee

Owing   to the properties   of the  spin variables,  the  values of the
correlation functions $G^{\alpha \beta}({\bf r}) $ at ${\bf r}={\bf 0}$ can be expressed
in terms of the magnetizations as
\be
\label{hardspin1}
G^{\alpha\beta}({\bf r=0})=-(1+m^\alpha)(1+m^\beta) \hspace{1.2cm}  \alpha \neq \beta 
\ee
\be
\label{hardspin2}
G^{\alpha\alpha}({\bf r=0})=-(1+m^\alpha)(1+m^\alpha-q). 
\ee
On  the  other hand, the enthalpy  of  the system  (which  is just the
internal  energy in the  absence  of  external  field) depends  on the
values of the correlation functions at n.n. separation. Using the
identity $\delta_{s_is_j}=\sum_{\alpha=1}^q \delta_{s_i \alpha}\delta_{s_j\alpha}$, we get
 
\be
\label{hen}
{\cal   E}=\frac{\partial\tilde     {\cal   G}}{    \partial\beta}=-    J   \sum_{<ij>}
<(q\delta_{s_i,s_j}-1)> = -\frac{NcJ} {2  q } \sum_{\alpha=1}^{q} [G^{\alpha\alpha}({\bf
r}={\bf e}) + (m^{\alpha})^2] \ ,
\ee
where  $c$ is the  coordination number of the   lattice, and ${\bf e}$
denotes a vector from the origin to one of its nearest neighbors.

For the theory to be thermodynamically consistent, the Gibbs potential
$\cal{G}$    must   have the   same   value   when  obtained  via  the
susceptibility route, i.e., Eq. (11),  or via the energy (or enthalpy)
route,  i.e.,   Eq.~(\ref{hen}). This  thermodynamic  consistency   is
embodied in  Maxwell relations  between the partial derivatives of
${\cal G}$ with respect to  the independent control variables. In  the
present case, these relations have the following form:
\be
\label{maxwell} 
 \frac {\partial} {\partial \lambda}  \widehat{C}^{\alpha\beta}({\bf k=0})=-\frac {1} {2q}\frac {\partial^2} {\partial
{m}^{ \alpha} \partial {m^{\beta}}  }  \sum_{\gamma=1}^{q}[G^{\gamma \gamma }({\bf r=e}) +
(m^{\gamma})^2 ]  \hspace{1.2cm}  \alpha,\beta=1,2,...,(q-1)\ ,
\ee
which  derive from Eqs.   (11) and (14)  (from now on the coordination
number  $c$ is absorbed   in  the inverse temperature variable  $\lambda=\beta
cJ$).  These   partial   differential  equations (PDE)     extend  the
self-consistent    equation  considered in   Ref.\cite{DS1996} for the
spin-$\frac{1}{2}$ Ising model to $(q-1)$ order parameters.

The  theory presented in this paper  is based on approximations for the
structure  of the direct  correlation functions  and  the above  exact
equations are the starting point of our analysis.

\subsection{Ornstein-Zernike approximation}

In general, the direct correlation functions are expected to remain of
finite range, even at a  critical point\cite{HMCDO1976}. Following the
Ornstein-Zernike approximation, we assume  here that they have exactly
the range of  the  exchange  interaction  in the  Potts   Hamiltonian,
Eq. (1),  which implies  that they  are truncated at  nearest-neighbor
separation.  We thus write
\be
\label{nnappro}
C^{\alpha\beta}({\bf r})=C_0^{\alpha\beta} \delta_{{\bf r},{\bf 0}}+ C_1^{\alpha\beta} \delta_{{\bf
r},{\bf e}} \ ,
\ee
or in Fourier space,
\be
\label{nnapprok}
\widehat{C}^{\alpha\beta}({\bf k})=C_0^{\alpha\beta}+c \ C_1^{\alpha\beta} \widehat{\lambda}({\bf k})
\ee
where  $\widehat{\lambda}({\bf  k})=\frac {1}   {c}  \sum_{{\bf  e}} e^{i{\bf
k}.{\bf  e}}$  is the  characteristic function  of   the lattice.  The
$q(q-1)$ distinct  functions $C_0^{\alpha\beta}$  and $C_1^{\alpha\beta}$ depend  on
the inverse temperature and on the magnetizations, but this dependence
is  not   given {\it  a priori}.   As  is   well-known in liquid-state
theory\cite{HMCDO1976}, when one assumes some approximate but explicit
dependence of the direct  correlation functions upon the thermodynamic
variables,  as    in  the  random   phase   approximation  or  in  the
mean-spherical   approximation,     the    theory    is   in   general
thermodynamically    inconsistent.       In  our    self-consistent OZ
approximation,  the   functions  $C_0^{\alpha\beta}$  and  $C_1^{\alpha\beta}$   are
determined     by    enforcing    the     zero-separation  conditions,
Eqs.~(\ref{hardspin1})-(\ref{hardspin2}),   and        by     imposing
thermodynamic consistency via Eq.  ~(\ref{maxwell}).

The above equations,  Eqs.  (10)-(17), define  the  SCOZA for  the Potts
model in the  most general situation, i.e., in  the presence of  $q-1$
independent ordering fields.  To make it a workable scheme, one has to
ensure that the number of equations equals  the number of unknowns and
that the PDE's are supplemented with proper boundary conditions.  This
is the  case for  $q=2$ (the  Ising   model), and more  generally  for
integer    values of  $q\geq     2$.   Indeed,   Eqs.~(\ref{hardspin1}),
(\ref{hardspin2}), and  (\ref{maxwell})  provide  $q(q-1)$ independent
equations for the $q(q-1)$ unknowns,  and the boundary conditions  for
the $q$-state  model may be  obtained  either exactly (for  $\lambda=0$ and
$m^{\alpha}=  q-1$) or from the solution  of the  $(q-1)$-state model (for
$m^{\alpha}=-1$).  Therefore,  when $q$ is  an integer, one can  solve the
equations  recursively.  In practice,   however,  this is a  difficult
numerical  problem, as one has  to solve $q(q-1)/2$ coupled PDE's with
$q$ independent variables,  the $q-1$  independent magnetizations plus
the inverse  temperature  (the equations  must be solved  in  the full
$q$-dimensional space even if   one is  interested ultimately  in  the
zero-field  case only).  It   is thus necessary  to  develop a simpler
approach  in  which   the  number  of  PDE's     does not  depend   on
$q$. Moreover, since our goal is to study the  Potts model for integer
as well as non-integer values of $q$, one should be able to perform an
analytic  continuation  of the  number  of states   to arbitrary  real
values.

\section{A SCOZA for non-integer values of $\lowercase{q}$}

\subsection{The partial differential equation}

To be able to study  non-integer as well as  integer values of $q$, we
consider the case of a magnetic field that acts only on one state, say
state 1,  while preserving the symmetry  between the remaining $(q-1)$
states.    Without  loss  of generality,  we    take  $h^1=h\neq  0$ and
$h^2=h^3=\ldots  h^q=0$.  It follows that $m^2=m^3=\ldots=m^q=-m^1/(q-1)$, and
there is only one independent  order parameter, $m^1=m$.  By symmetry,
one has also only four distinct connected (resp.  direct ) correlation
functions, for instance   $G^{11}$, $G^{22}$, $G^{12}$,  and  $G^{23}$
(resp.  $C^{11}$, $C^{22}$, $C^{12}$, and $C^{23}$).   It is then easy
to block-diagonalize the   matrices  of correlation  functions   so to
rewrite the OZ equations, Eq.~(\ref{OZK}), as
\be
\label{OZsym}
\left(
\begin{array}{cc}
\widehat{G}^{11}({\bf k}) & \widehat{G}^{12}({\bf k})  \\
(q-2)\widehat{G}^{12}({\bf      k})       &     [\widehat{G}^{22}({\bf
k})+(q-3)\widehat{G}^{23}({\bf k})] \\
\end{array}
\right)
\left(
\begin{array}{cc}
\widehat{C}^{11}({\bf k}) & \widehat{C}^{12}({\bf k}) \\
(q-2)\widehat{C}^{12}({\bf      k})     &       [\widehat{C}^{22}({\bf
k})+(q-3)\widehat{C}^{23}({\bf k})] \\
\end{array}
\right)=1
\ee

\be
\label{OZasym}
[\widehat{G}^{22}({\bf                       k})-\widehat{G}^{23}({\bf
k})][\widehat{C}^{22}({\bf k})-\widehat{C}^{23}({\bf k})]=1.
\ee
Moreover, Eq. (6) implies that
\be
\label{G12}
G^{12}=-\frac {G^{11}} {q-1}
\ee
\be
\label{G23}
G^{23}=\frac {G^{11}} {(q-1)(q-2)}-\frac {G^{22}} {q-2} \ ,  
\ee
from  which   we   obtain,  by replacing   in   Eqs.~(\ref{OZsym}) and
(\ref{OZasym}),

\be
\label{C12}
C^{12}=C^{23}=\frac{1}{2}C^{22}.
\ee
We finally get
\be
\label{G11}
\widehat{G}^{11}({\bf k})=\frac {1} {\widehat{C}^{11}({\bf k})-\frac{1}{2}\frac{q-2}
{q-1}\widehat{C}^{22}({\bf k})} \ ,
\ee
and
\be
\label{G22}
\widehat{G}^{22}({\bf k})=\frac{\widehat{G}^{11}({\bf k})}{(q-1)^2} +2
\ \frac {q-2}{q-1}\frac{1}{\widehat{C}^{22}({\bf k})}  \ .
\ee
The enthalpy ${\cal E}$ (see Eq.~(\ref{hen})) is now given by
\be
\label{hens}
{\cal   E}=-\frac     {NcJ} {2q}  [G^{11}({\bf  r=e})+(q-1)G^{22}({\bf
r=e})+\frac {q}{q-1}m^2],
\ee
and  since   there is only  one  order  parameter, there  is  only one
inverse-susceptibility relation,
\be
\label{dgdm2}
\frac {\partial^2 {\tilde{\cal G}}/N} {\partial m^2}=\frac{\partial {\tilde h}}{\partial m}=\widehat{C}^{11}({\bf k}={\bf 0})-\frac{1}{2}\frac {q-2} {q-1}\widehat{C}^{22}({\bf k}={\bf 0})
 \ .
\ee
The partial differential equation expressing thermodynamic consistency
follows from Eq.~(\ref{maxwell}):
\bea
\label{maxwells}
\frac {\partial }{\partial \lambda}  [\widehat{C}^{11}({\bf k}={\bf 0})-\frac{1}{2}\frac {q-2} {q-1}\widehat{C}^{22}({\bf k}={\bf 0})
] & = & -\frac {1} {(q-1)}  \\ \nonumber
  & & - \frac {1} {2q} \frac {\partial^2}
{\partial m^2} [G^{11}({\bf
r=e})+(q-1)G^{22}({\bf r=e})].   
\eea
So    far,    all   equations are     exact.   

The OZ   approximation,  Eq.~(\ref{nnapprok}),  introduces  4  unknown
functions, $C_0^{11}$, $C_0^{22}$, $C_1^{11}$, $C_1^{22}$,
\bea
\label{approxks}
\widehat{C}^{11}({\bf k})=C_0^{11}+c\ C_1^{11} \widehat{\lambda}({\bf k}) \\ 
\widehat{C}^{22}({\bf k})=C_0^{22}+c\  C_1^{22} \widehat{\lambda}({\bf k})
\ .
\eea
Introducing the variables 
\be
\label{z1}
z= -c \frac {C_1^{11}-\frac{1}{2}\frac{q-2}{q-1} C_1^{22}}
{C_0^{11}-\frac{1}{2}\frac{q-2}{q-1} C_0^{22}}
\ee
and
\be
\label{z2}
z^{\star}=-c\frac {C_1^{22}}{C_0^{22}} \ ,
\ee
we have
\be
\label{C11z}
\widehat{C}^{11}({\bf k})-\frac{1}{2}\frac{q-2}
{q-1}\widehat{C}^{22}({\bf k})=[C_0^{11}-\frac{1}{2}\frac{q-2}
{q-1}C_0^{22}][1-z\widehat{\lambda}({\bf k})]
\ee
and
\be
\label{C22z}
\widehat{C}^{22}({\bf k})=C_0^{22}[1-z^{\star} \widehat{\lambda}({\bf k})] \ .
\ee
Using Eqs.~(\ref{G11},\ref{G22}) and  introducing the  lattice Green's
function\cite{J1971}
 
\be
\label{green}
P(z,{\bf r})= \frac{1}{(2\pi)^3} \int_{-\pi}^{\pi} d^3{\bf k}
\frac{e^{-i{\bf k}.{\bf r}}}{1-z\widehat{\lambda}({\bf k})} \ ,
\ee
we finally    obtain  the expressions    of   two-point connected
correlation functions in real space,
\be
\label{g11z}
G^{11}({\bf r})=\frac{P(z,{\bf r})}{C_0^{11}-\frac{1}{2} \frac{q-2}
{q-1}C_0^{22}}
\ee
and
\be
\label{g22z}
G^{22}({\bf r})=\frac{G^{11}({\bf r})}{(q-1)^2}+2 \frac
{q-2}{q-1}\frac{P(z^{\star},{\bf r})}{C_0^{22}} \ .
\ee
The two unknown quantities, $C_0^{11}$ and $C_0^{22}$, are  eliminated by
enforcing the exact zero-separation conditions following from
Eqs.~(12) and (13), namely
\be
\label{g110}
G^{11}({\bf r=0})=-(1+m)(1+m-q),
\ee
\be
\label{g220}
G^{22}({\bf r=0})=-\frac {[(q-1)^2+m][1+m-q]}{(q-1)^2} \ .
\ee
This yields

\be
\label{g11}
G^{11}({\bf r})=(1+m)(q-1-m)\frac{P(z,{\bf r})}{P(z)}
\ee
\be
\label{g22}
G^{22}({\bf r})=\frac {G^{11}({\bf r})} {(q-1)^2}
+\frac{q(q-2)(q-1-m)} {(q-1)^2} \frac{P(z^{\star},{\bf r})}{P(z^{\star})} \ ,
\ee
where $P(z)\equiv P(z,{\bf r}={\bf 0})$.  By inserting  the above results  in
Eq.~(\ref{maxwells}), the following  PDE is obtained:
\bea
\label{eqscoza}
\frac {\partial }{\partial \lambda} [P(z)(1-z)] & = & - \frac {(1+m)(q-1-m)}
{(q-1)} \\ \nonumber 
 & & ( 1+ \frac{1}{2} \frac {\partial^2} {\partial
m^2}\{ [q-1-m][(1+m)\frac{P(z)-1}{zP(z)}+(q-2)\frac{P(z^*)-1}{z^*P(z^*)}] \} )
\eea
where the relation $P(z,{\bf r}={\bf e})=(P(z)-1)/z$ has been used. We
see that $q$ is now a parameter that can take any real value.  Setting
$q=2$  in Eq.~(\ref{eqscoza}) gives back  the  PDE for the Ising model
with only one unknown function, $z(\lambda,m)$\cite{DS1996}. In the general
case, however, there     are  two unknown   functions,   $z(\lambda,m)$ and
$z^*(\lambda,m)$, but only one equation relating them.  Additional input is
therefore needed.   We  address this  point as  well  as the  required
boundary conditions for Eq.~(\ref{eqscoza}) in the following.

\subsection{Boundary conditions and further approximation}

The full range of variation  of $\lambda$ is from  $0$ to $+\infty$ and that of
the magnetization $m$ goes from $-1$ to $(q-1)$. The initial condition
for Eq.~(\ref{eqscoza}) at $\lambda=0$ is provided by the exact solution of
the noninteracting  Potts model  in  an external  field.  One  has  in
particular       $C_1^{11}(\lambda=0,m)=C_1^{22}(\lambda    =0,m)=0$, so    that
$z(\lambda=0,m)=z^{\star}(\lambda  =0,m)=0$.  Similarly,  the condition for $m=q-1$
follows from the fact that all site variables are in state $1$ ($h
\to  \infty  $),  which  again  leads to  $z(\lambda,m=q-1)=z^{\star}(\lambda ,m=q-1)=0$
because $C_0^{11}$  and $C_0^{22}$ $\to +\infty$.   At the boundary $m=-1$,
the state   $1$ is  suppressed:   the  system then  corresponds  to  a
$(q-1)$-state Potts model in zero field,  whose solution is unknown in
general. However, in order  to study the  zero-field transition of the
$q$-state Potts model ($q>1$) that leads to preferential occupation of
state   $1$, it  is  sufficient  to  consider the  region $m\geq0$.  The
boundary condition   at $m=0$ results  from the  $Z_q$ symmetry of the
Hamiltonian in Eq.~(\ref{Hamilt})  when the external fields are turned
off: above the transition  temperature,  one must have  $h(\lambda ,m=0)=0$
and all   $q$ states  must   be  equivalent.  As  a consequence,   the
correlation functions $G^{11}$  and  $G^{22}$, and similarly  $C^{11}$
and  $C^{22}$, are  equal; hence, $C_0^{11}(\lambda ,m=0)=C_0^{22}(\lambda,m=0)$
and    $C_1^{11}(\lambda   ,m=0)=C_1^{22}(\lambda,m=0)$,    which   implies  via
Eqs.~(\ref{z1}) and  (\ref{z2}) that  $z(\lambda,m=0)=z^{\star}(\lambda,m=0)$.  The
condition $h(\lambda,m=0)=0$ can  be reexpressed as a constraint
on the enthalpy ${\cal  E}$. Indeed, since  the boundary condition  at
$\lambda=0$    garantees that   $h(\lambda=0,m=0)=0$,   and since  thermodynamic
consistency  implies that  $\partial  {\tilde  h}/ \partial \lambda\arrowvert_{
m}=1/(NcJ)  \   \partial {\cal  E}/\partial m\arrowvert_{   \lambda  }$, the zero-field
condition in the disordered phase requires   that $ \partial  {\cal E}/ \partial
m\arrowvert_{ \lambda,m=0 }=0$.

Having settled  the boundary-condition problem,  we must still find an
additional equation that, together with Eq.~(\ref{eqscoza}), allows to
uniquely determine the  functions $z(\lambda, m)$  and $z^{\star}(\lambda , m)$. We
do so by  considering the exact  expressions of the direct correlation
functions at nearest  neighbor separation, $C_1^{11}$ and  $C_1^{22}$,
at  the  two  boundaries $m=0$    and $m=q-1$.   We already  saw  that
$C_1^{11}(\lambda,m=0)=C_1^{22}(\lambda,m=0)$  as    a  result  of     the $Z_q$
symmetry. The calculation  for $m=q-1$ is detailed in Appendix  A
and leads to the following expression:
\be
\label{DCp-1}
\Delta C(\lambda)\equiv C_1^{11}(\lambda,m=q-1)-C_1^{22}(\lambda,m=q-1)=\frac {1} {q^2}
(e^{\lambda q/c}-1)^2.
\ee
We then make a linear interpolation between the $m=0$ and $m=q-1$
values for all temperatures:
\be
\label{H3}
 C_1^{11}(\lambda,m)-C_1^{22}(\lambda,m) \simeq \frac {m} {(q-1)} \Delta
C(\lambda) \ ,
\ee
which yields the following additional equation for $z(\lambda,m)$ and $z^*(\lambda,m)$:
\be
\label{H3z}
zP(z)-(1+m)z^{\star}P(z^{\star}) = -c(1+m)(q-1-m)\frac{m}{(q-1)}\Delta C(\lambda),
\ee
where we have used Eqs.~(\ref{g11z})-(\ref{g220}) to eliminate
$C_0^{11}$ and $C_0^{22}$.

\subsection{Numerical procedure}

The numerical   integration   of  Eq.~(\ref{eqscoza}),  coupled   with
Eq. (\ref{H3z}), is carried out by an explicit algorithm, the
partial   derivatives   being    approximated by    finite  difference
representations. Eq.~(\ref{eqscoza})   is  rewritten  as  a  nonlinear
diffusionlike equation,
\be
\label{eqscoza2}
\frac{\partial z}{\partial \lambda} =-\frac
{(1+m)(q-1-m)}{(q-1)[(1-z)P'(z)-P(z)]}
(1+\frac{1}{2}\frac {\partial^2 \Delta {\cal E}}{\partial m^2}  )
\ee
with             $P'(z)\equiv         dP(z)/dz$         and            $\Delta
{\cal{E}}=(q-1-m)\{(1+m)[P(z)-1]/[zP(z)]+(q-2)[P(z^*)-1]/[z^*P(z^*)]\}$.      The
derivative of $z$ with respect to $\lambda$ is obtained from
\be
\label{dJ}
\left.\frac {\partial z} {\partial \lambda}\right|_{m}=\frac {z(\lambda +\delta \lambda,m) -z(\lambda,m) }
{\delta \lambda} \ ,
\ee
where $\delta  \lambda$ is the elementary step  in  the ``time-like'' direction,
and the second derivative  of $\Delta  {\cal E}$  with  respect to $m$  is
given by
\be
\label{dm2}
\left.\frac{\partial^2 \Delta {\cal{E}}} {\partial m^2}\right|_{\lambda}=
\frac{ {\Delta \cal E}(\lambda,m+\delta m)-2{\Delta \cal E}(\lambda,m)+{\Delta \cal
E}(\lambda,m-\delta m)} {{\delta 
m}^2} \ ,
\ee
where $\delta m$ is the elementary step in the ``space-like'' direction.

One   starts from   $\lambda=0$    for  which  $z=z^*=\Delta   {\cal   E}=0$
(noninteracting system). Then,  Eq.~(\ref{eqscoza2}) gives  $z$ at the
next step  in  the $\lambda$-direction, $\lambda=\delta  \lambda$.   At this temperature,
$z^{\star}$ is  calculated   from  Eq.~(\ref{H3z})  by a   Newton-Raphson
algorithm.   Once $z$ and  $z^{\star}$ are  known, $\Delta {\cal  E}$ and its
second derivative are computed.  At the next  step, the temperature is
decreased ($\lambda \to \lambda +\delta \lambda$) and the same procedure is repeated.  $\delta
\lambda$ is  gradually decreased as  the spinodal is approached (see below)
to  ensure numerical stability of  the explicit scheme.  As previously
mentioned, the boundary condition at   $m=q-1$ is known exactly,  with
$z(m=q-1)=z^{\star}(m=q-1)=0$, and the  zero-field condition at  $m=0$ is
enforced by setting $\Delta {\cal E}\arrowvert_{ m=\delta m}=\Delta {\cal E}
\arrowvert_{ m=- \delta m}$ which readily yields from Eq. (47) $\left. \partial^2 \Delta
{\cal  E}/ \partial  m^2  \right|_{m=0}=   2[\Delta {\cal E}(\lambda,\delta   m)-\Delta {\cal
E}(\lambda,0)]/ \delta m^2$

It follows from Eqs.(26) and (32) that the inverse susceptibility goes
to zero when  $z\to 1$.  (Note that from  the definition of the lattice
Green's function, Eq.~(\ref{green}), $z$ and $z^{\star}$ must stay in the
interval $[0,1]$.)  The locus of the points for  which $z=1$ defines a
spinodal curve in the $m-T$ plane.  We denote $T_c$ the temperature at
which the spinodal has its maximum.  When  the maximum of the spinodal
occurs   at $m=0$, the   zero-field  transition is  continuous and the
spontaneous  magnetization, obtained   from  the condition  $h(m,T)=\partial
({\cal G}/N)/ \partial  m=0$, varies with $T$  as depicted  schematically in
Fig.  (1a).  When  the maximum of the spinodal  occurs at $m=m_c\neq 0$,
the   zero-field   transition  is  first-order   and  the  spontaneous
magnetization  jumps  to a non-zero  value  $\Delta  m$  at  a  transition
temperature $T_t<T_c$, as shown in Fig.  (1b).  The transition remains
first-order in the presence of a small non-zero external field and for
$h=h_c$ it ends   in a second  order  critical point. This   critical
endpoint corresponds to the maximum of the spinodal.

In the numerical    procedure, the spinodal  curve  is  never  reached
exactly.   With the  choice  of  a   small parameter  $\epsilon$  (typically
$\epsilon=10^{-5}$ or $10^{-6}$), we  define a pseudo-spinodal as the points
of the   grid formed by  the discretized  values of  $m$ and $\lambda$ that
satisfy $1<z<1-\epsilon$. The    pseudo-spinodal then serves as  a  boundary
condition for   the finite-difference equation  in the low-temperature
phase.   In this  phase,  $\delta  \lambda$ is  fixed  to   a very small  value
(typically $10^{-6}$ or $10^{-7}$).  We have checked that the location
of the spinodal does  not move when $\epsilon$ is  changed from $10^{-5}$ to
$10^{-6}$  and that the  whole calculation is stable  when  $\delta \lambda$ is
decreased from $10^{-6}$ to $10^{-7}$. All  calculations are done with
$\delta m=(q-1)/200$.

The numerical integration has  been carried out  for the sc, bcc,  and
fcc lattices.   The  corresponding Green's  functions  $P(z)$ are well
documented\cite{J1971}  and can be    easily tabulated.  To   test the
numerical procedure, we  have checked that  the results  for the Ising
model are  recovered with good  accuracy.  For instance, we have found
$\beta_c  J=0.22126$  for  the inverse critical   temperature   on the sc
lattice,    in perfect  agreement     with  the   value obtained    in
Ref.~\cite{DS1996}.

\section{Results and discussion}
\label{sec:results}

\subsection{$q=3$ and $4$}

We have first considered the $3$ and $4$-state Potts models that have
been much studied in the  recent literature (the $3$-state model plays
a important role in both condensed matter and high-energy physics, and
the $4$-state model is a special case of the Ashkin-Teller model).  On
the three lattices, we have clearly  found that the system undergoes a
first-order transition with a critical point in non-zero magnetization
(or  field).  For illustration,  we show  in Fig.  2  the spinodal and
spontaneous  magnetization curves for the   sc lattice.  The (inverse)
transition temperatures, given in Table I,  are in very good agreement
with the   most   recent  estimates from   Monte    carlo simulations,
$q\beta_tJ=0.550   565(10)$    for  the  $3$-state   model     on the  sc
lattice\cite{JV97} and $q\beta_tJ=0.628 616(16)$  and $0.455 008(10)$ for
the $4$-state   model       on   the    sc  and       bcc    lattices,
respectively\cite{AZ97}. (To the best of our knowledge, the transition
temperatures for the fcc  lattice are given  here for the first time.)
Remarkably, it appears that the precision of the theory ($0.05 \%$) is
better  than for the Ising model\cite{DS1996},  despite the use of the
additional approximation  for $C_1^{11}  -C_1^{22}$  which has  led to
Eq. (44). This  is probably due to   the fact the correlation  length,
although large, remains finite at the transition.

We also give in Table  I the predictions  for the dimensionless latent
heat $\Delta e=e(T_t^+)  -e(T_t^-)$,  where $e=(c/2) <\delta _{s_is_j}>$,  and
for  the jump discontinuity in  the  order parameter at the transition
$\Delta  m/(q-1)$ (with our   definition  of the local magnetizations   in
Eq. (3),  $m$ varies here between $0$  and $q-1$).  For  the $3$-state
model on  the sc lattice,  the predicted value of  the  latent heat is
somewhat  larger than the estimate  from  Monte Carlo simulations, $\Delta
e=0.1606 \pm 0.0006$\cite{ABV1991}, but it  is smaller than the estimate
from series expansions, $\Delta e=0.264 \pm  0.011$\cite{GE94}. (As noted in
Ref.\cite{GE94},  the large difference between  these two estimates is
surprising and has  received no explanation.)  Similarly, our value of
$\Delta m/2$ is in the range defined by the Monte Carlo estimate, $\Delta m/2=
0.395\pm 0.005 $\cite{GKP89},  and  the series expansion  estimate,  $\Delta
m/2=0.505$\cite{GE94}.   Our  results also show that   $\Delta  m$ is only
weakly dependent on the lattice structure.  This quasi-universality is
a  consequence of the  large value  of  the correlation length  at the
transition.  (In two dimensions,  $\Delta m$ is the  same for the  square,
triangular  and honeycomb lattices, but this   is a consequence of the
star-triangle relation\cite{B1982}.)

For completeness, we  also report in Table   I the coordinates  of the
critical endpoint in non-zero external field.  For the $3$-state model
on  the  sc lattice,  the theoretical  predictions  are  very close to
recent    Monte    Carlo      estimates,    $q\beta_cJ=0.54938(2)$    and
$\beta_ch_c=0.000775(10)$\cite{KS2000}.  We may  note,  however, that the
accuracy for the critical temperature is ``only'' $0.4\%$.

\subsection{$2<q<3$}

As q is decreased below $3$, $\Delta m$ and $h_c$ decrease and $\beta_c$ gets
close to $\beta_t$, as shown in Table II  in the case  of the sc lattice.
This    indicates that the  first-order    character of the zero-field
transition becomes  weaker and weaker.  For  $q<2.4$,  the spinodal is
extremely   flat and  to within the   accuracy  of our calculations we
cannot distinguish $\beta_t$ from $\beta_c$ nor $h_c$ (or $\Delta m$) from zero.
Therefore, the direct analysis of the numerical results cannot yield a
clear-cut conclusion   regarding the order of   the transition in this
range  of $q$.   On the other  hand,   the asymptotic analysis  of the
equations in the vicinity of a  putative critical point indicates that
because of the linear interpolation formula for the direct correlation
functions   at  n.n.    separation,  Eq.   (43),    the transition  is
first-order  for all   $q>2$: indeed,  as  shown  in Appendix  B, this
additional approximation is incompatible with the zero-field condition
$ \partial   {\cal E}/ \partial m\arrowvert_{  \lambda,m=0  }=0$ for  $q>2$  if  the
transition is continuous. This  latter conclusion is  likely to  be an
artifact   of the too  crude   linear interpolation formula. A  proper
resolution of the question of the existence of a critical value $q_c>2$
would  require a fully self-consistent SCOZA  that does not rely on an
approximate,  explicit relation between  $z$ and $z^*$. In the absence
of such a theory, interesting insight into the potential change of the
transition from first- to second-order can still be gained by studying
the  numerical results  in  the region of   $q$ where the  first-order
character of the transition  can be clearly identified with  numerical
precision, i.e., for    $q\geq 2.4$. As   illustrated by  the very  good
accuracy  of the theoretical predictions  for  $q=3$,  we expect  this
region to be less dependent on the detail of  the relation between $z$
and $z^*$, and the  putative changeover behavior  at $q_c$ can then be
investigated by extrapolation of the data to $q$ less than $2.4$.

This procedure has been used to analyze the data shown in Figs.  3 and
4 for the latent heat $\Delta  e$ and the second-moment correlation length
$\xi$ in the disordered phase at the  transition point ($\xi$ is defined
by $\widehat{G}^{11}({\bf k})  \sim\widehat{G}^{11}({\bf 0})(1+\xi^2k^2),
k\to 0$; specifically, one has $c\xi^2  =z/(1-z)$).  For the sc lattice,
we find that $\Delta e$ is well fitted in the range $2.4\leq q\leq 2.8$ by the
simple  algebraic form $\Delta e  \sim a(q-q_{c})^{\sigma}$  with $q_{c}\approx 2.15$
and $\sigma\approx  1.70$ (if the result for  $q=3$ is also taken into account,
we find instead $q_c\approx 2.21$ and $ \sigma \approx 1.44$).  Similarly, the rapid
increase of  $\xi$ is well  described by  $\xi \sim b(q-q_{c})^{-\nu}$ with
$q_c\approx 2.14$ and $\nu\approx 1.14$ (this fit is not modified when the result
for $q=3$ is added, as can be  seen in Fig.  4).   This value of $q_c$
compares  favorably  with  the  estimate  obtained via   a  real space
renormalization group  method\cite{NRS1981} ($q_c\approx  2.2$),  a  method
that yields the excellent  value $q_c=4.08$ in  two dimensions;  it is
smaller    than   the     estimates   obtained   from   $1/q$   series
expansions\cite{KS82} ($q_{c}=2.57\pm  0.12$ with  $\sigma=1.65\pm 0.15$ ) and
from   Monte-Carlo  simulations\cite{LK91}  ($q_{c}=2.45\pm  0.1$   with
$\nu=0.87\pm 0.08$).  (Note however that  the estimate of Ref.\cite{LK91}
is obtained  by extrapolating numerical data in  the  range $2.8\leq q\leq
3.2$ .) In the  case of the bcc  lattice, the same value $q_c\approx  2.14$
(to  within our numerical accuracy) is   obtained from analysis of the
latent heat and the correlation length, with exponents $\sigma\approx 1.61$ and
$\nu  \approx 1.11$, respectively.  On  the other hand,  no reliable results
are  obtained for the fcc  lattice. This is due  probably  to the fact
that the asymptotic regime is not yet  reached in the range $2.4\leq q\leq
2.8$, as can be seen in Fig. 4 with the small values of the
correlation length $\xi$.

Further  investigation of changeover   behavior can be  carried out by
considering the  $q$-dependence   of the coordinates   of the critical
endpoint  (restricting again the numerical  study  to $q\geq2.4$).  When
$q\to    q_{c}$,   one  expects  that $h_c    \to    0$ and   $T_c(q) \to
T_{cc}=T_c(q_c)=T_t(q_c)$. As shown in Fig.   5, the  data for the  sc
lattice are again  well   fitted by the simple   form $T_{cc}-T_c(q)\sim
h_c(q)^{\rho}$ with $k_BT_{cc}/(Jc)=0.775\pm  0.003$ and  $\rho=0.39\pm 0.02$.
(Note  that the same  power-law behavior with  the mean-field exponent
$\rho=0.4=2/5$ is observed when a  tricritical point is approached along
a critical wing line\cite{LS1984}.)  As can be   seen in Fig. 6,  this
result fits in with the global $T-q$ phase diagram of the model on the
sc lattice (this figure  also includes the results  for $q\leq  2.4$ for
which  no difference  is  found  between  $T_t$ and  $T_c$  within the
accuracy  of our calculations).   For  $T=T_{cc}$, we now find  $q_c\approx
2.15$, which is consistent with the preceding estimates.

\subsection{$1<q\leq 2$ and the bond-percolation limit}

When $q\leq 2$, the transition  is clearly continuous, as illustrated in
Fig. 7 for $q=2$ and $q=1.8$.  One can thus investigate the asymptotic
behavior of  the  zero-field susceptibility as  the  critical point is
approached from the  high-temperature  phase.  Fig.  8 shows   log-log
plots of the reduced susceptibility $\chi_{red}(q)=k_BT\chi(q)/(q-1)$ as a
function of the reduced temperature $t=1-T_c(q)/T$ (the factor ($q-1$)
comes  again from our definition of  the magnetization).  We also plot
in the figure the results for $q=2.07$ for which the susceptibility is
apparently  diverging, despite  the fact that   the transition is  not
really continuous, as discussed above and  in Appendix B. In the whole
range $1<q\leq 2$, the divergence  of $\chi(q)$ is governed asymptotically
by the critical exponent $\gamma=2$ which  is that of the spherical model.
However, the results are affected by a  strong crossover, and for $t\geq
10^{-2}$  the divergence  of  $\chi(q)$  is   governed by  an  effective
exponent that increases as $q$  decreases (see Table III).  For $q=2$,
this exponent is close to the exact  Ising value $\gamma\approx 1.24$, as noted
in earlier work\cite{DS1996}.  Note that  the increase of $\gamma$ as  $q$
decreases is the exact behavior of the two-dimensional model (see e.g,
Ref.\cite{W82}).

An interesting  limit  is $q\to 1^+$,   for which there  is  a well known
mapping  to  the bond  percolation  problem\cite{KF1969}: the critical
temperature of the Potts model  is related to the percolation treshold
via the  equation  $p_c/(1-p_c)=e^{J\beta_c(q=1)}-1$  and  $\gamma$  becomes  the
exponent  for  the  mean cluster  size.   For $q=1.001$,  we find
$p_c=0.241, 0.172$ and   $0.114$ for the   sc, bcc, and  fcc lattices,
respectively.  These   values are in   reasonable agreement  with  the
current  best estimates $p_c=0.2488, 0.1803$ and $0.119$\cite{SA1994}.
Moreover, in the range  $t\geq 10^{-2}$, the divergence of the
susceptiblity is governed by the effective  exponent  $\gamma\approx 1.71$,
which is  close to the expected value for three dimensional lattices,
$\gamma\approx 1.80$.

\section{Conclusion}

We have  applied a thermodynamically  self-consistent OZ approximation
to the ferromagnetic   Potts  model on  the various  three-dimensional
lattices..   The present theory  gives an accurate  description of the
phase diagram of  the model for  all values of $q>1$.   The transition
temperatures for   instance  are within  $0.05\%$   of the best  known
estimates  for $q=3$ and $q=4$   (first-order transitions) and even in
the   limit $q\to  1^+$ where   the  transition  temperature  gives the
percolation threshold  of the    associated bond  percolation  problem
(second-order transition), they  are within $4\%$  of the  Monte Carlo
estimates  for  the sc, bcc,   and  fcc lattices.    The SCOZA has  an
intrinsic limitation: the OZ   assumption restricting the range of  the
direct correlation functions to that of the pair interaction potential
prevents an accurate  treatment of  the asymptotic critical  behavior.
However,   the     region  of   concern appears   to   be  very  small
$(|T-T_c|/T_c<10^{-2})$,  which   allows  the  theory     to make   good
predictions  for   the effective  critical  exponents.  One  finds for
instance $\gamma \approx 1.28$  instead of $1.24$  for $q=2$ (Ising model)  and
$\gamma \approx 1.71$  instead of $1.80$  for $q\to 1^+$ (percolation). 

 Another shortcoming  that we have   encountered in the present  study
 results from the following fact: in order to describe the Potts model
 for  arbitrary   real values   of  $q$ and  thereby   investigate the
 potential change in the order of the transition,  one has to restrict
 the parameter space  to a subspace  preserving the $Z_{q-1}$ symmetry
 between all but one   of the $q$ states.    As a consequence of  this
 restriction,  the   SCOZA  approach becomes   incomplete,  with  more
 unknowns than   exact conditions    to  be satisfied.      Additional
 approximations  must    thus be  devised   that  go  beyond  the mere
 application of the OZ formalism.   The approximation we chose in this
 work (see Eq.  (43)) still allows,  as stressed above, to locate with
 high accuracy the  transition   temperatures and to provide   a  good
 description of the thermodynamic quantities over  a wide range of the
 parameters $(T,m,q)$.   It also gives   numerical indication that the
 order of the transition seems to change from second to first order at
 a  $q_c$  slightly  above $2$.   Yet, a   proper  description of this
 changeover phenomenon would require  a better approximation than that
 used here. Making progress on this issue, that also plagues the SCOZA
 approach for systems with  quenched disorder in the replica formalism
 (because   of the necessity   to   make an analytic continuation  for
 letting the number of replicas go  to zero\cite{KRT97}), is certainly
 an important problem  that  must still be   addressed to confirm  the
 SCOZA as an accurate and a versatile theoretical scheme.

\acknowledgments

We are grateful to E. Kierlik for useful discussions  and for his help
in the numerical computations.

\newpage

\appendix

\section{}

In   this appendix, we derive the    exact expressions of $C_1^{11}$ and
$C_2^{22}$ for $m=q-1$.  These expressions  are used to build the 
approximate relationship between $z$ and $z^*$, Eq. (44).

When $m=q-1$, all  spins are in state  $1$ and $\tilde   h=\beta h \to \infty$.  In
order  to get an expansion in  powers of $u=(q-1-m)$,   we consider, at a
given temperature, the  excited states corresponding to  configurations
where one, two, or more spins are in a  state different from $1$.  The
partition  function is  first  expressed in  powers of  $y=e^{- \tilde
h}$.   To derive Eq.~(\ref{DCp-1}),  we  only need to consider the
first and the second  excited states.  At   this order, the  partition
function $Z(\beta J,\tilde h)$ in any dimension depends only on the
coordination number  $c$ of the lattice. We find
\bea
\label{partition}
Z(\beta J,\tilde h) & = & x^{\frac{-Nc(q-1)}{2}}y^{-N(q-1)}
[
1+N(q-1)x^{qc}y^q+\frac{N(N-1-c)(q-1)^2}{2} x^{2qc}y^{2q} \\ \nonumber+
 & & \frac{Nc(q-1)}{2} x^{2q(c-1)}y^{2q} 
 + \frac{Nc(q-1)(q-2)}{2} x^{q(2c-1)}y^{2q}+O(y^{3q})]
\eea
with $x=e^{- \beta J}$. 
We deduce from Eq.~(\ref{partition}) the expansion
for the  magnetization,
\bea
\label{magn}
m & =&\frac{1}{N}\frac {\partial \ln(Z)}{\partial \tilde
h}=  q-1-q(q-1)x^{qc}y^{q} \\ \nonumber 
 & & +q(q-1)x^{2qc}[(q-1)(1+c)+x^{-q}(2-q)c-cx^{-2q}]y^{2q}+
O(y^{3q}) \ .
\eea
The inversion of Eq.~(\ref{magn}) gives the expansion of $y^{q}$ in powers of $u$,
\be
\label{yp}
y^{q}=\frac{1}{q(q-1)x^{qc}}u+\frac{[(q-1)(1+c)+x^{-q}(2-q)c-cx^{-2q}]}{q^2(q-1)^2x^{qc}}u^2+
O(u^3).
\ee
The expansion of the susceptibility $\chi=\partial m /\partial \tilde h$  then follows,
\be
\label{chi}
 \chi = qu-\frac{[(q-1)(1+c)+x^{-q}(2-q)c-cx^{-2q}]}{q-1} u^2+ O(u^3).
\ee
Using $    k_BT \chi=     G^{11}({\bf k=0})=(q-1-m)(1+m)[1+z+ O(z^2)]$
(see Eq. (39)), we then find
\be
\label{zu}
z=\frac {c[x^{-2q}+(q-2)x^{-q}-(q-1)]}{q(q-1)}u+ O(u^2).
\ee

To derive Eq.~(\ref{DCp-1}), the expansion of $z^{\star}$ is also needed.
We then expand  the enthalpy ${\cal{E}}$, Eq.~(\ref{hens}), in powers
of $u$,
\be
\label{henu}
\frac{{\cal E}}{NcJ}=\frac{-1}{2(q-1)}[(q-1)^2-2(q-1)u+(1+\frac{qz_1}{c}
+\frac{(q-2)z_1^{\star}}{c})u^2+ O(u^3)] \ ,
\ee
where $z=z_1u+ O(u^2)$, $z^*=z^*_1u+ O(u^2)$. Note that in writing Eq.
(A6) as in deriving Eq. (A5), we have used the fact that the SCOZA
structure with the direct correlation functions extending only to
n.n. separation is exact at the order in $u$ considered. Comparing Eq.
(A6) to the expansion  of ${\cal{E}}=-J\  \partial \ln(Z)/\partial (\beta J)\arrowvert_{{\tilde h}}$ that is derived from
Eqs.~(\ref{partition}) and (\ref{yp}), 
\be
\label{henu2}
\frac{{\cal
E}}{NcJ}=-\frac{(q-1)}{2}+u-\frac{1}{q(q-1)}[x^{-2q}+x^{-q}\frac{q-2}{2}]u^2+
O(u^3) \ ,
\ee
we find
\be
\label{zstaru}
z^{\star}=\frac{c}{q(q-1)}x^{-q}[x^{-q}-1]u+ O(u^2) \ .
\ee
Since, from Eqs. (30,31), 
\be
C_1^{11}=-\frac{1}{cq}[z_1+(q-2)z^*_1]+O(u)
\ee
and
\be
C_1^{22}=-\frac{2(q-1)}{cq}z^*_1+O(u) \ , 
\ee
we finally obtain

\be
C_1^{11}(\lambda,m=q-1)-C_1^{22}(\lambda,m=q-1)=\frac{1}{q^2}(1-e^{\lambda q/c})^2.
\ee

\newpage

\section{}

In  this  appendix, we  show  that  the  proposed  theory cannot truly
predict  a second-order transition  (in zero external field) for $q>2$
because of the approximation   given in Eq.  (43), that   represents a
linear interpolation formula for $C_1^{11}-C_1^{22}$.

Let  us assume  that the  transition is  continuous. Then, the inverse
susceptibility $\chi^{-1}=\partial h/ \partial m$ for $m=0$ should go to zero at the
transition. This implies  that $z\to 1$ when  $m\to  0$ at  the critical
temperature, and we may expand all  quantities in the vicinity of $z=1$
and $m=0$. Since  the $3$-dimensional lattice Green's function behaves
as
\be
\label{B1}
P(z)\sim P(1)[1-a\epsilon+ O(\epsilon^2)],
\ee
where        $\epsilon=\sqrt{1-z}$    and       $a$      is    a    positive
constant\cite{J1971}, we get from Eqs. (25), (39) and (40) that  the
enthalpy  behaves  as
\be
{\cal E}\sim\frac{P(1)-1}{P(1)}[1-\frac{a}{(q-1)P(1)}(\epsilon+(q-2)\epsilon^*)]+O(m^2,\epsilon^2,{\epsilon^*}^2)
\ee
 where $\epsilon^*=\sqrt{1-z^*}$. 

As discussed in section IIIB, the zero-field condition at $m=0$ in the
disordered phase implies that $  \partial {\cal E}/ \partial m\arrowvert_{  \lambda,m=0
}=0$. From the above expression, this implies that

\be
\left.\frac{\partial ( \epsilon+(q-2)\epsilon^*)}{\partial m}\right|_{\lambda,m=0}=0 \ .
\ee
We must thus have
\be
\epsilon+(q-2)\epsilon^*=O(m^{1+u})
\ee
with  $u>0$. 

On the other hand, expanding  Eq. (44) near the critical point yields
\be
\label{B2}
\epsilon^*-\epsilon\sim \frac{P(1)-c\Delta C(\lambda_c)}{aP(1)}m +O(m^2),
\ee
where $\Delta C(\lambda_c)$, given by Eq.~(\ref{DCp-1}), is positive.

There are two possibilities for Eqs. (B4) and (B5) to be compatible:

$(1)$  Neither $\epsilon$ nor  $\epsilon^{\star}$ have a term  linear in $m$, thereby
implying via Eq.(B5) that $P(1)-c\Delta  C(\lambda_c)=0$. The numerical  study,
however, shows that for $2<q<2.4$  this quantity is small but strictly
positive  in  the  range of  temperatures where   the transition takes
place.

$(2)$ Both $\epsilon$ and $\epsilon^{\star}$ have terms linear  in $m$, with Eq. (B5)
satisfied (for  $q\neq2$), and there is  an  exact cancellation of these
terms so   that Eq.  (B4)  is  also obeyed.    However, since $\epsilon$ and
$\epsilon^{\star}$ are  both non-negative  quantities, this cancellation cannot
occur for $q>2$.

Therefore, there cannot  be a zero-field continuous transition for $q>2$.

\newpage

\begin{table} 
\caption[99]{Theoretical  predictions for the inverse transition
temperature, the dimensionless  latent heat, the jump discontinuity in
the  order parameter, and the  coordinates of the critical endpoint on
the sc, bcc and fcc lattices for the $3$ and $4$-state Potts models.}

\begin{tabular}{ccccccccc}
%  & \multicolumn{3}{c}{$q=3$} & \multicolumn{3}{c}{$q=4$}   \\
  & lattice & $q\beta_tJ$ & $\Delta e$ & $\Delta m/(q-1)$ & $q\beta_cJ$
&$\beta_ch_c$ \\
\tableline
q=3 &  sc  &	0.5507	& 0.199& 0.444 & 0.5474 & 0.00080\\
    & bcc   &	0.3957	& 0.319 & 0.452 & 0.3926 & 0.00135\\
    & fcc  &	0.2576	& 0.519 & 0.457 & 0.2556 & 0.00124\\
\tableline
q=4 & sc  & 0.6288 & 0.576& 0.628& 0.6134 & 0.00491\\
    & bcc & 0.4549 & 0.870& 0.636 & 0.4418 & 0.00622\\
    & fcc & 0.2970 & 1.371& 0.638 & 0.2880 & 0.00670\\
\end{tabular}

\label{tab:1}
\end{table} 

\newpage

\begin{table} 
\caption[99]{Theoretical predictions for the inverse transition temperature, the dimensionless latent heat, the jump discontinuity
in  the order parameter, and  the coordinates of the critical endpoint
on the simple cubic lattice for $2.4\leq q\leq 2.8$ }

\begin{tabular}{ccccccccc}
  
& $q$  &   $q\beta_tJ$ & $\Delta e$ & $\Delta m/(q-1)$ & $q\beta_cJ$  &$\beta_ch_c$ \\
\tableline
& 2.8  &  0.5321 & 0.132 & 0.387 & 0.5302 &  0.00039\\
& 2.7  &  0.5223 & 0.099 & 0.349 & 0.5209 &  0.00025\\
& 2.6  &  0.5122 & 0.072 & 0.314 & 0.5113 &	0.00014\\
& 2.5  &  0.5017 & 0.045 & 0.265 & 0.5011 &  0.00007 \\
& 2.4  &  0.4908 & 0.027 & 0.226 & 0.4904 &	0.00003\\
\end{tabular}

\label{tab:2}
\end{table} 

\newpage

\begin{table}    
\caption[99]{Theoretical  predictions for the effective critical
exponent $\gamma_{eff}(q)$ that governs the divergence of the
susceptibility for $(T-T_c)/T\geq 10^{-2}$.}

\begin{tabular}{cccc}
  
& $q$  &   $\gamma_{eff}$ & \\
\tableline
& 2 &  1.28 & \\
& 1.9&  1.34 & \\
& 1.8 &  1.39 & \\
& 1.7&  1.45 & \\
& 1.001& 1.71 &\\
\end{tabular}

\label{tab:3}
\end{table}

\newpage

\begin{figure}
\begin{center}
\epsfxsize= 440pt
\epsffile{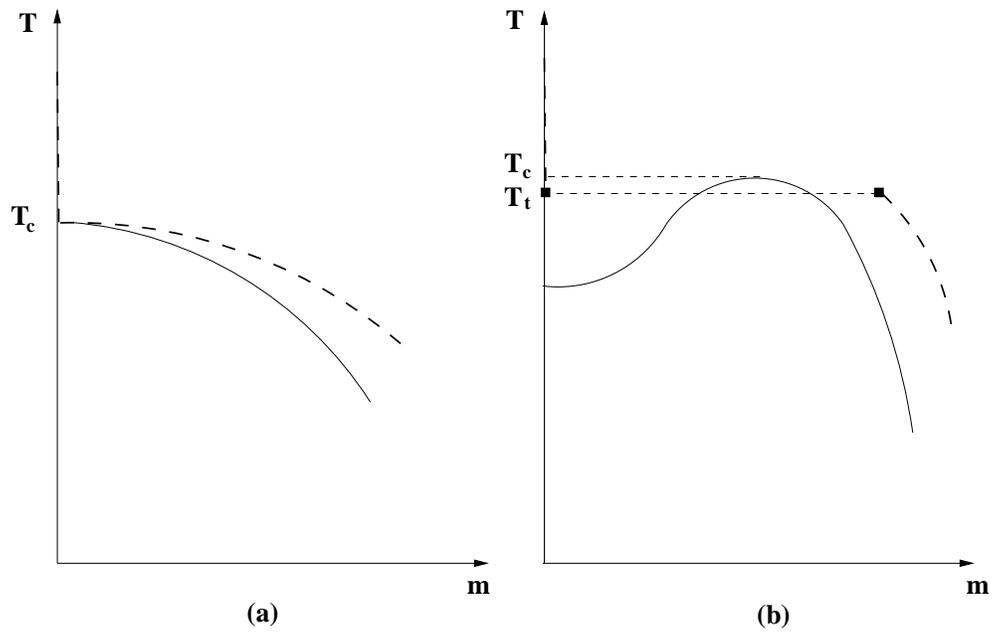}
\vspace{1cm}
\caption{\label{fig1} Schematic plot of the spinodal curve (solid line) and the spontaneous 
magnetization versus temperature (dashed line) for second-order (a) and 
first-order (b) transitions. }
\end{center}
\end{figure}

\newpage

\begin{figure}
\begin{center}
\epsfxsize= 450pt
\epsffile{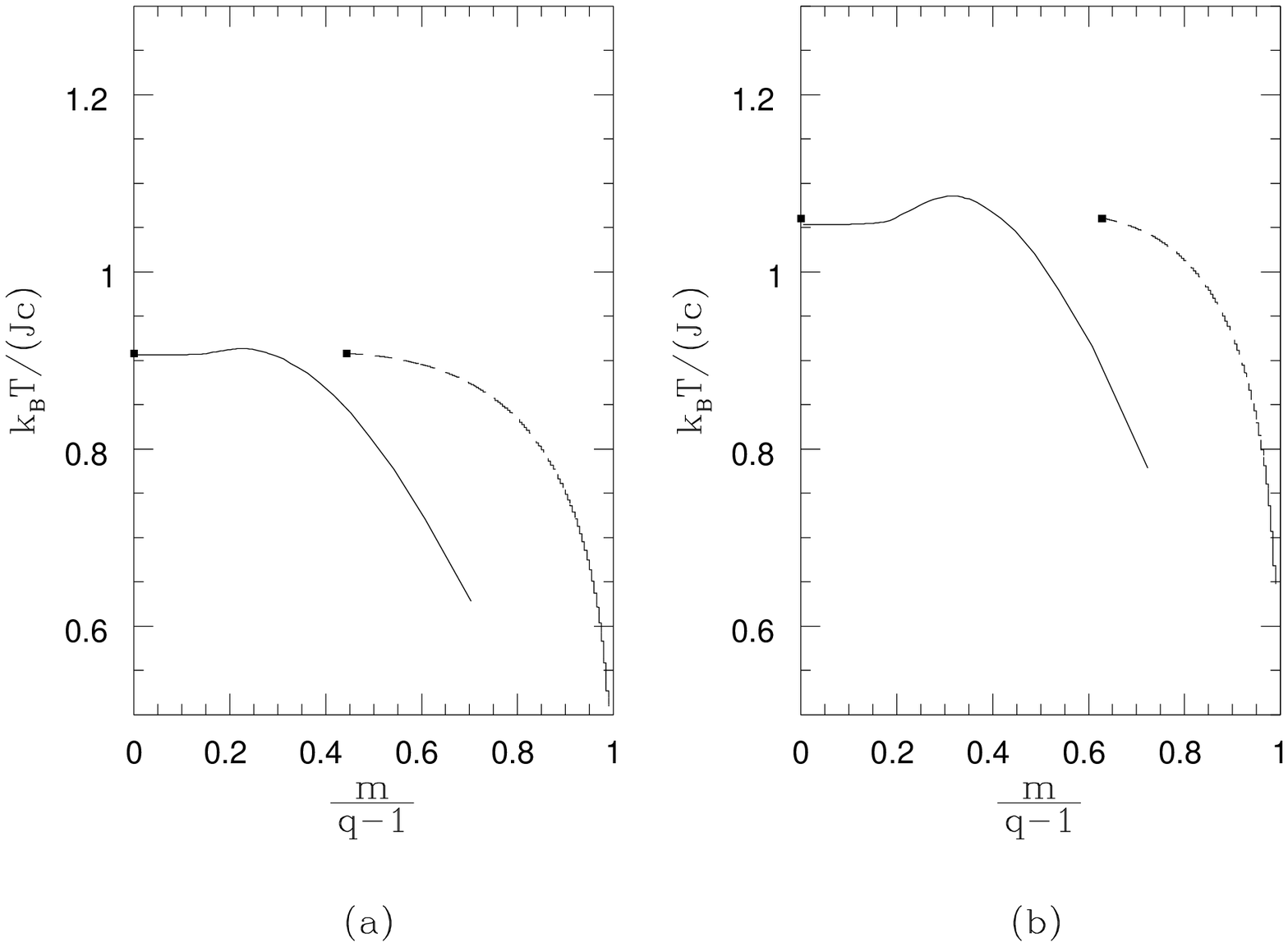}
\vspace{1cm}
\caption{\label{fig2} Spinodal  curve  (solid line)   and  spontaneous magnetization
versus temperature (dashed line) for $q=3$ (a) and $q=4$ (b). }
\end{center}
\end{figure}

\newpage

\begin{figure}
\begin{center}
\epsfxsize= 350pt
\epsffile{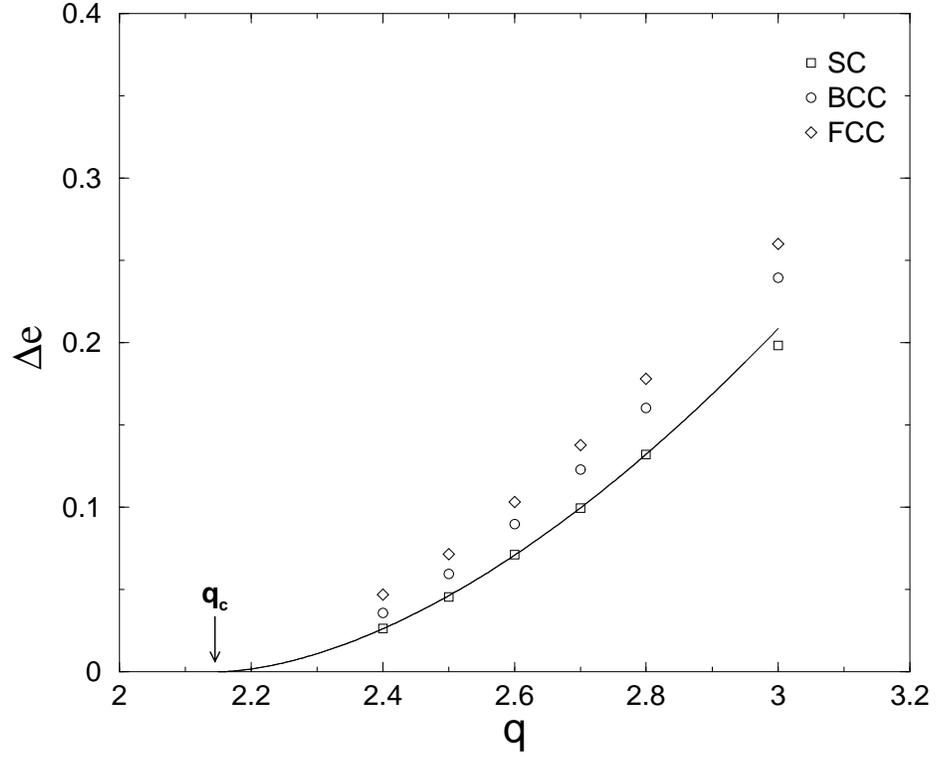}
\vspace{1cm}
\caption{\label{fig3} Latent heat $\Delta e$ on the sc, bcc, and fcc lattices for
$2.4\leq q\leq 3$.  The solid line is the  algebraic fit, $\Delta e \sim  0.275 \
(q-2.15)^{1.70} $, of the data for the sc lattice.}
\end{center}
\end{figure}

\newpage

\begin{figure}
\begin{center}
\epsfxsize= 350pt
\epsffile{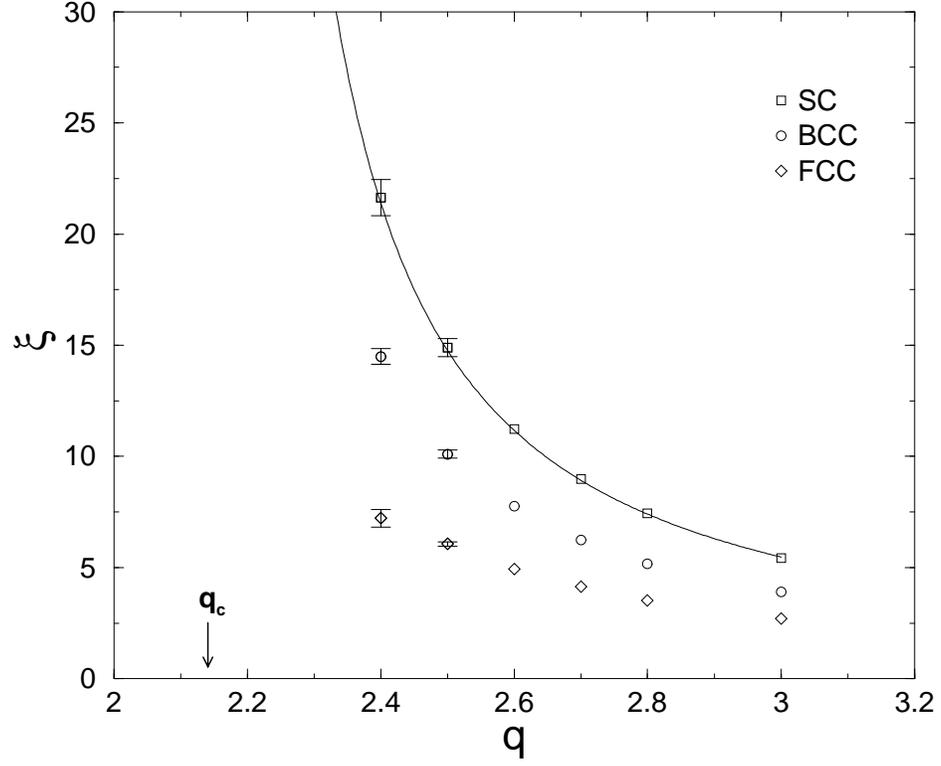}
\vspace{1cm}
\caption{\label{fig4} Second-moment  correlation length  on the   sc, bcc, and  fcc
lattices for $2.4\leq q\leq 3$.  The solid  line is the algebraic fit, $\xi
\sim 4.61 \ (q-2.14)^{1.14}$, of the data for the sc lattice}
\end{center}
\end{figure}

\newpage

\begin{figure}
\begin{center}
\epsfxsize= 450pt
\epsffile{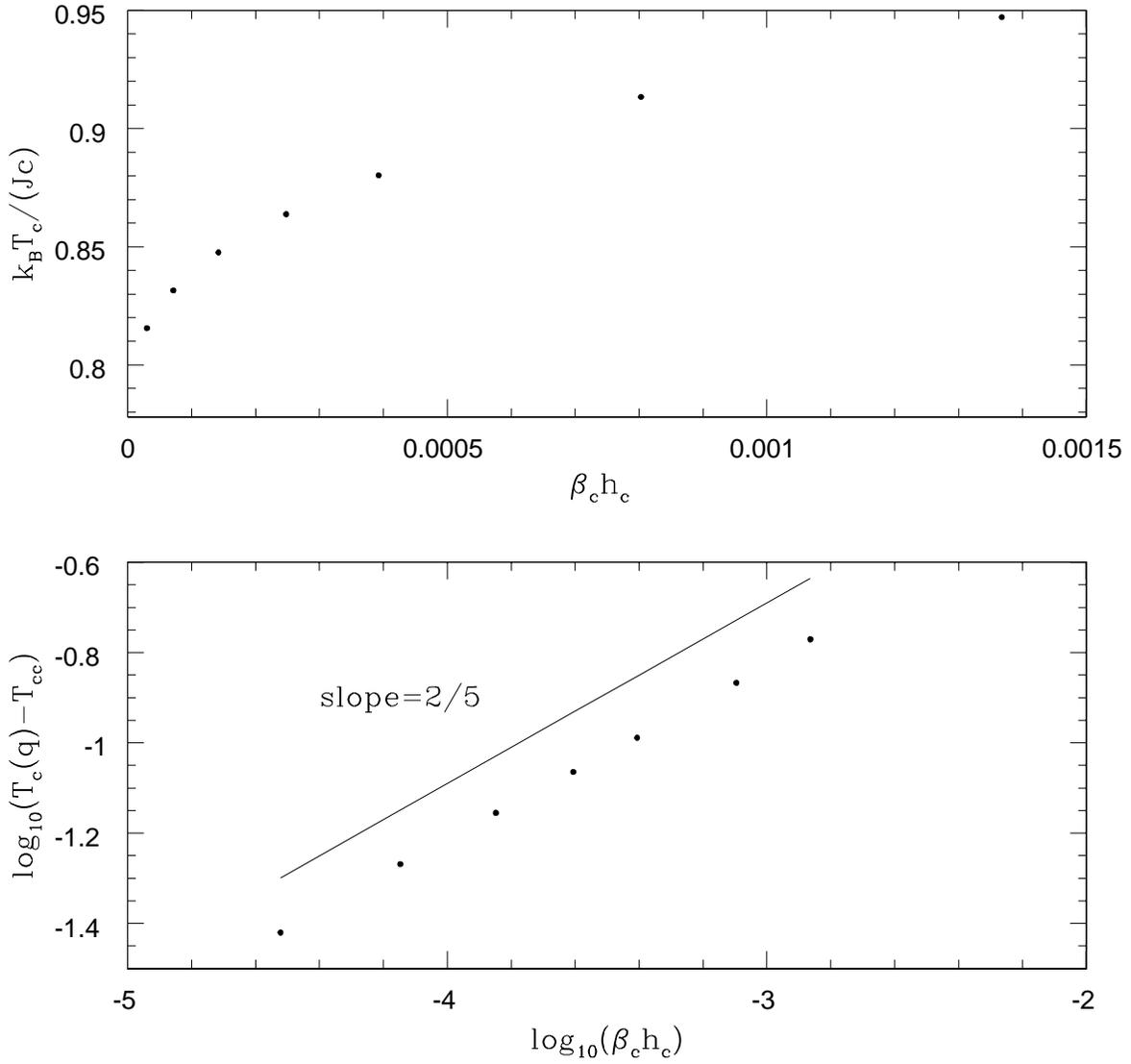}
\vspace{1cm}
\caption{\label{fig5} Coordinates of the critical endpoint  in the  $T-h$ plane for
$q=2.4,  2.5, 2.6, 2.7, 2.8, 3$,  and $3.2$ (from  left to right). The
corresponding  log-log   plot  indicates    that  $h_c$  vanishes   at
$k_BT_{cc}/(Jc)=0.775\pm0.003$.}
\end{center}
\end{figure}

\newpage

\begin{figure}
\begin{center}
\epsfxsize= 350pt
\epsffile{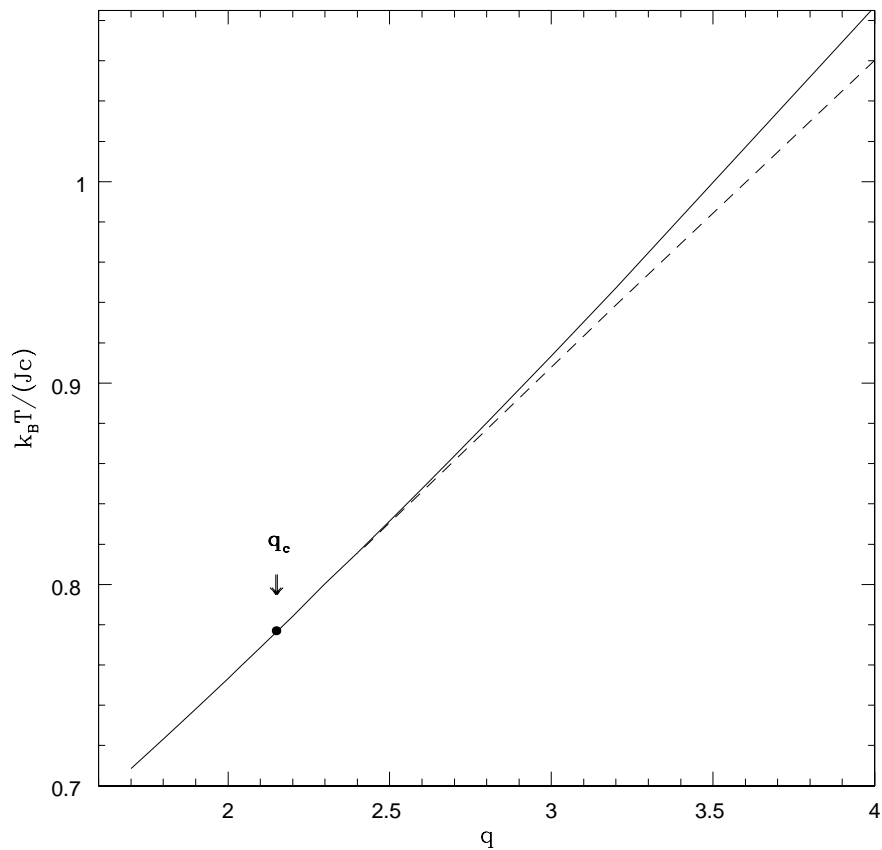}
\vspace{1cm}
\caption{\label{fig6} Predicted phase diagram of the $q$-state Potts model on the sc
lattice.  The solid and dashed   curves represent the second-order and
first-order lines, respectively. The second-order transition occurs in
zero field only for $q<q_c$. (Note that within numerical accuracy, our
results cannot distinguish first from second-order for $2<q<2.4$.)}
\end{center}
\end{figure}

\newpage

\begin{figure}
\begin{center}
\epsfxsize= 450pt
\epsffile{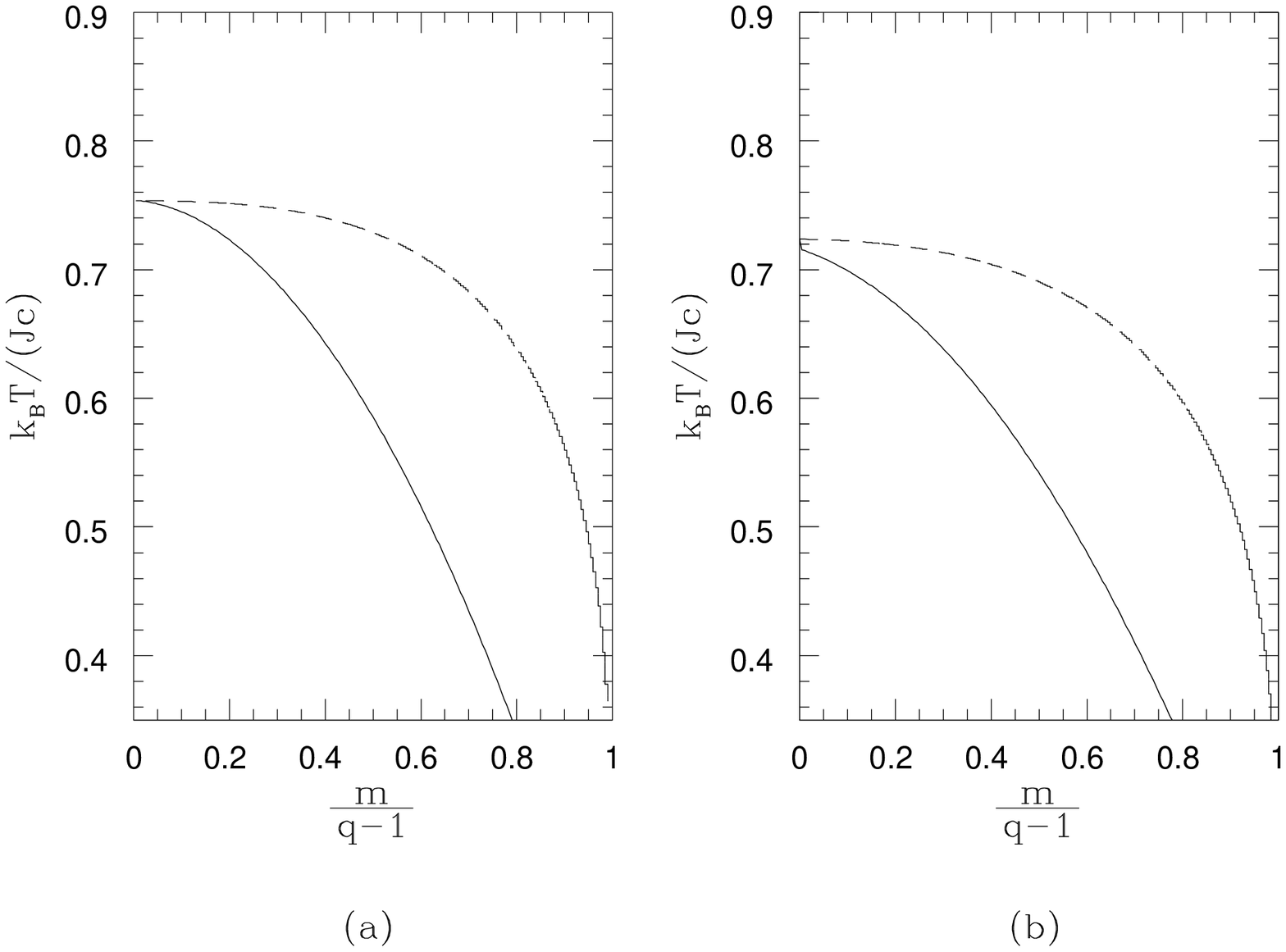}
\vspace{1cm}
\caption{\label{fig7}Spinodal  curve  (solid line)   and  spontaneous magnetization
versus temperature (dashed line) for $q=2$ (a) and $q=1.8$ (b).}
\end{center}
\end{figure}

\newpage

\begin{figure}
\begin{center}
\epsfxsize= 350pt
\epsffile{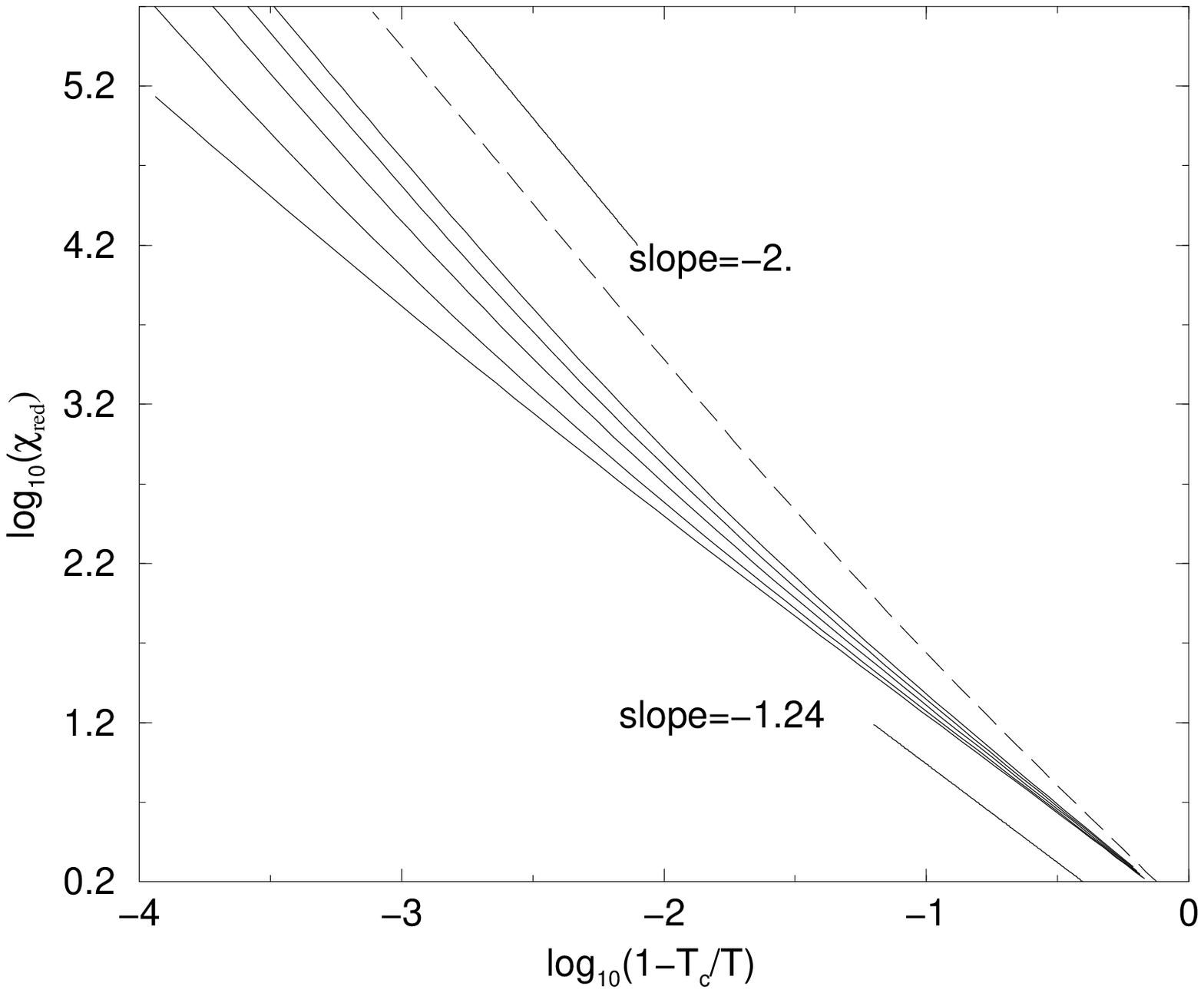}
\vspace{1cm}
\caption{\label{fig8} Log-log  plot    of the  reduced zero-field   susceptibility
$\chi_{red}=k_BT\chi/(q-1)$ versus the reduced temperature $t=1-T_c/T$ for
$q=2.07, 2, 1.9, 1.8, 1.7$ (solid lines from bottom to top)
and $q=1.001$ (dashed line).}
\end{center}
\end{figure}


\begin{references}
\bibitem[*]{AAAuth} The Laboratoire de Physique Th{\'e}orique des Liquides is the UMR 7600 of the CNRS.
\bibitem{P52}
R.~B. Potts, Proc. Camb. Phil. Soc. {\bf 48} (1952) 106.

\bibitem{W82}
F.~Y. Wu, Rev. Mod. Phys. {\bf 54} (1982)  235; J. Appl. Phys. {\bf
55} (1984) 2421.

\bibitem{K54}
T. Kihara, Y. Midzuno, and J. Shizume, J. Phys. Soc. Jpn. {\bf 9}
(1954) 681.


\bibitem{Ba73}
R.~J Baxter, J. Phys. C. {\bf 6} (1973) L445.

\bibitem{JV97}
W. Janke and R. Villanova, Nucl. Phys. B {\bf B489}  (1997) 679   and
references therein.

\bibitem{NRS1981}
B. Nienhuis, E.~K. Riedel, and M. Schick, Phys.  Rev. B {\bf 23}
(1981)  6055.

\bibitem{KS82}
J.~B. Kogut and D.~K. Sinclair, Solid State Communications {\bf 41}
(1982) 187.

\bibitem{LK91}
J. Lee and J.~M. Kosterlitz, Phys. Rev. B {\bf 43} (1991)  1268.


\bibitem{HS77} J. S. Hoye and G. Stell, J. Chem. Phys. {\bf 67} (1977) 439; Mol. Phys. {\bf 52} (1984)  1071; Int. J. Thermophys. {\bf 6} (1985)
561.

\bibitem{DS1996} R. Dickman and G. Stell, Phys. Rev. Lett. {\bf 77} (1996)
996; D. Pini, G. Stell, and R. Dickman, Phys. Rev. E {\bf
57} (1998)  2862.

\bibitem{GKRT2000}
S. Grollau,  E. Kierlik, M.~L. Rosinberg,  and G. Tarjus, 
Phys. Rev. E., (in press).

\bibitem{KRT97}
E. Kierlik, M.~L. Rosinberg, and G. Tarjus, J. Stat. Phys. {\bf 89} (1997)  215; {\bf 94} (1999)  805; {\bf 100} (2000)  423.

\bibitem{HMCDO1976}
J.~P. Hansen and I.~R. McDonald, Theory of Simple Liquids
(Academic, New York, 1976).

\bibitem{J1971}
G.~S. Joyce, J. of Math. Phys. {\bf 12} (1971)  1390; J. Phys. C {\bf
4} (1971)  L53; J. Phys. A. {\bf 5} (1972) L65.


\bibitem{AZ97}
P. Arnold and Y. Zhang, Nucl. Phys. B {\bf 501} (1997)  803.

\bibitem{ABV1991}
N.~A. Alves, B.~A Berg, and R. Villanova, Phys. Rev. B {\bf 43} (1991)
5846.

\bibitem{GE94}
A.~J. Guttmann and G. Enting, J. Phys. A: Math. Gen. {\bf 27} (1994)  5801.

\bibitem{GKP89}
R.~V. Gavai, F. Karsh, and B. Petersson, Nucl. Phys. B {\bf 322} (1989)  738.

\bibitem{B1982}
R.~J Baxter, J. Phys. A.: Math. Gen.  {\bf 15} (1982) 3329.

\bibitem{KS2000}
F. Karsch and S. Stickan, Phys. Lett. B {\bf 488} (2000) 319.

\bibitem{LS1984}I. D. Lawrie and S. Sarbach in: Phase transitions
and  Critical   Phenomena, C.     Domb and J.  L.  Lebowitz Eds. 
(Academic, London, 1984), Vol. 9.


\bibitem{KF1969}
P.W. Kasteleyn and E.M. Fortuin, J. Phys. Soc.  Japn. Suppl. {\bf 26} (1969)
11; Physica {\bf 57} (1972)  536.

\bibitem{SA1994}
D. Stauffer and A. Aharony, Introduction to Percolation Theory
(Taylor \& Francis, London, 1994).



\end{references}
\end{document}